\documentclass{PoS}

\usepackage{textcomp} 

\newcommand{\veritas}{\textit{VERITAS}}
\newcommand{\fermilat}{\textit{Fermi}-LAT}
\newcommand{\fermi}{\textit{Fermi}}
\newcommand{\swiftuvot}{\textit{Swift}-UVOT}
\newcommand{\uvot}{\textit{UVOT}}
\newcommand{\swiftxrt}{\textit{Swift}-XRT}
\newcommand{\xrt}{\textit{XRT}}
\newcommand{\swift}{\textit{Swift}}
\newcommand{\magic}{\textit{MAGIC}}
\newcommand{\flwo}{\textit{FLWO 48"}}
\newcommand{\hess}{\textit{H.E.S.S.}}

\newcommand{\onees}{\textit{1ES~1727+502}}

\title{\veritas\ detection of $\gamma$-ray flaring activity from the BL Lac object \onees\ during bright moonlight observations}

\ShortTitle{\veritas\ detection of $\gamma$-ray flaring activity from the BL Lac object \onees\ }

\author{\speaker{Matteo Cerruti}\ for the \veritas\ Collaboration\\
        Harvard-Smithsonian Center for Astrophysics, 60 Garden St., Cambridge, MA 02138, USA\\
        E-mail: \email{matteo.cerruti@cfa.harvard.edu}}

\abstract{During May 2013, a gamma-ray flare from the BL Lac object \onees\ (z=0.055) has been detected with the \veritas\ Cherenkov telescopes. This detection represents the first evidence of very-high-energy (E>100 GeV) variability from this blazar and has been achieved using a reduced-high-voltage configuration which allows observations under bright moonlight. The integral flux is about five times higher than the archival VHE flux measured by MAGIC. The detection triggered additional \veritas\ observations during standard dark-time and multiwavelength observations from infrared to X-rays with the \flwo\ telescope and the Swift satellite. The results from this campaign are presented and used to produce the first spectral energy distribution of this object during gamma-ray flaring activity. The spectral energy distribution is then fit with a standard synchrotron-self-Compton model, placing constraints on the properties of the emitting region in the blazar.}

\FullConference{The 34th International Cosmic Ray Conference,\\
		30 July- 6 August, 2015\\
		The Hague, The Netherlands}

\begin{document}

\section{Introduction}
\label{section1}

Active galactic nuclei (AGN) of the blazar class represent the most common source of very-high-energy (VHE; E > 100 GeV) $\gamma$-ray photons in the extra-galactic sky. Their emission is characterized by a broad spectral energy distribution (SED) from radio to $\gamma$-rays, extreme variability and a high degree of polarization. In the framework of the unified AGN model they are considered AGN whose relativistic jet is aligned with the line of sight.
The study of VHE blazars is complicated by their broadband emission and their rapid variability. Strictly simultaneous multiwavelength (MWL) campaigns are the key to study blazar physics.\\

The cameras of imaging atmospheric Cherenkov telescopes are made of photomultiplier tubes (PMTs), which cannot safely operate in bright moonlight conditions. This constraint particularly affects the studies of blazars (and other variable sources), limiting the organization of MWL campaigns, or prohibiting follow-up observations of a flare.  Among the current generation of IACTs, only \magic\  and \veritas\ \cite{Magicmoon00, Rico07, Magicmoon} observe under moderate moonlight ($< 35\%$ moon illumination).  \textit{FACT}, whose camera is instead composed of solid-state Geiger-mode avalanche photodiodes (also called silicon-photomultipliers) \cite{FACT2013}, is also capable of performing moonlight observations.\\

In 2012, the \veritas\ collaboration began a new observing program under bright moonlight by reducing the high voltage (RHV) applied to the PMTs, or by utilizing UV-transparent filters. The details of the observing strategy are discussed in \cite{SeanICRC}. In this proceeding we present the detection of VHE flaring activity from the blazar \onees\ during May 2013, at a flux of roughly five times the archival VHE flux measured by \magic\ \cite{MAGICpaper}. The high-flux state was initially detected during bright moonlight observations, which represents an innovation for \veritas.\\

 The blazar \onees\  is a nearby high-frequency-peaked BL Lac object ($z = 0.055$), discovered as a $\gamma$-ray source by \fermi\ and as a VHE source by \magic\ \cite{MAGICpaper}. The \magic\ collaboration reported an integral flux of $(5.5 \pm 1.4)\times 10^{-12}$ cm$^{-2}$ s$^{-1}$ above $150$ GeV and a spectral index of $2.7 \pm 0.5$ \cite{MAGICpaper}.\\

Further details on the \veritas\ multi-wavelength campaign are presented in \cite{1727apj}.

\section{\veritas\ instrument} 
\label{section2}

\veritas\ is an array of four IACTs located at the Whipple Observatory in southern Arizona at an altitude of 1.3~km above sea level. Each telescope is of Davies-Cotton design with a 12-m diameter reflector, and the array is arranged in a diamond with $\sim$ 100~m to a side.  Each \veritas\ telescope is instrumented with a camera made up of 499 PMTs each, with a total field-of-view of 3.5\textdegree. \veritas\ is able to reconstruct $\gamma$-ray emission from about 85 GeV to $>$30 TeV, and a single-event angular resolution of 0.1\textdegree\ at 1~TeV. During standard observations (i.e. when the moon is $< 35\%$ illuminated) a source with an integrated flux of 1\% of the Crab Nebula flux can be detected at the $5\sigma$ level in $\sim$~25~hours, and a 5\% Crab source in less than 2 hours. More information on the \veritas\ array can be found in \cite{Holder06, NaheeICRC}.\\

In the RHV observation mode, the PMT voltages are reduced to 81\% of their standard values during dark-sky observations, allowing \veritas\ to operate when the moon is 35-65\% illuminated. The sensitivity of the RHV observation mode is similar to the standard \veritas\ sensitivity, albeit with a higher energy threshold ($\sim$~200~GeV). The cuts used in this RHV analysis would allow for a 5\% Crab source to be detected in less than 2 hours.\\

\section{\veritas\ observations of \onees\ }
\label{section3}

 The \veritas\ observations that allowed for the detection of \onees\ were taken between May 1, 2013 and May 7, 2013. There were additional data taken on May 18, 2013 that did not result in a detection. 
 After quality selection, approximately 6 hours of data remain. Of these, 3 hours were taken in RHV mode on the first two nights of the exposure.  All observations were made in ``wobble'' mode, wherein the telescopes were pointed 0.5\textdegree\ away from the target to allow simultaneous measurements of the target and background regions. All data in the 2013 dataset were taken at elevations between 65$^\circ$ and 70$^\circ$. \\

\begin{figure*}[t!]
\begin{center}

\includegraphics[width=350pt]{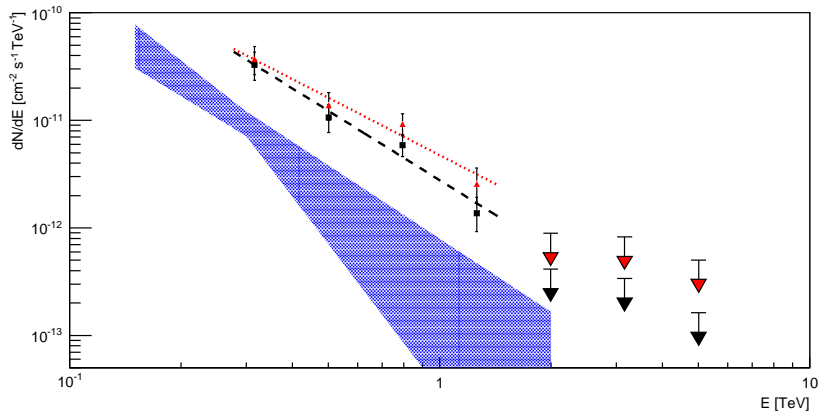}
\caption{Spectrum for the complete \onees\ dataset. Power-law fits to both the uncorrected (black) and EBL-corrected (red) points are provided. The blue shaded region represents the result from \magic\ [3]. For further details see \cite{1727apj}. }
\label{fig:veritasSpectrum}
\end{center}
\end{figure*}

The analysis of the complete 2013 dataset used the ``reflected-region'' 
background model \cite{2001A&A...370..112A} resulting in 159 ON events 
and 850 OFF events with a background normalization factor $\alpha$ of 0.077, 
yielding a detection significance of $9.3\sigma$ \cite{1983ApJ...272..317L}.
The reconstructed VHE spectrum is shown in Fig.~\ref{fig:veritasSpectrum}. 
In the same figure we show data points corrected for absorption by extragalactic background light (EBL) using the model in \cite{Franceschini08} (for z=0.055). 
Data are well fitted with a power-law function between 0.25 TeV and 1.6~TeV; the $\chi^2 / NDF$  is $1.66/2$ ($P \sim$  44\%). The observed spectrum is given by
 $ \frac{dN}{dE} = (7.8 \pm 1.1 )\times 10^{-12}  \left(\frac{E}{620~\mathrm{GeV}}\right)^{-2.1 \pm 0.3} \mathrm{cm^{-2}s^{-1}TeV^{-1}}$
All errors in the aforementioned fits are statistical only. The cumulative systematic errors on the flux normalization and spectral index are conservatively estimated to be 30\% and $\pm0.3$, respectively.\\

The light curve for the \veritas\ observations is shown in the top panel of Fig.~\ref{1727LC}. 
The peak integrated flux above 250~GeV is $(1.6 \pm 0.4)\times10^{-11}\mathrm{cm^{-2}s^{-1}}$, which corresponds to 9.5\% of the Crab Nebula flux. 
The upper limit is at the 95\% confidence level and represents $9.4\times10^{-12}\mathrm{cm^{-2}s^{-1}}$ (5.6\%~Crab). 
The light curve between May~01 and May~07 (MJD 56413-56419) can be fitted with a constant resulting in a flux of $(1.1\pm0.2)\times10^{-11}\mathrm{cm^{-2}s^{-1}}$ (6.3\%~Crab) and a $\chi^2/NDF = 4.73/3$ ($P \sim$ 19\%). \\

The integral flux above 250 GeV corresponds to about five times the flux measured by \magic, and represents the first evidence of VHE variability in \onees. The \veritas\ observations between May~01 and May~07 are consistent with a constant flux. However, when including the measurement on May~18 in the fit, the resulting $\chi^2/NDF$ is $12.0/4$ ($P \sim$ 1.8\%). Thus, a constant flux is excluded at the 2.4$\sigma$ level, indicating that the flare may have ended at some point after the last \veritas\ detection on May~07.\\

\section{Multiwavelength observations of \onees\ }
\label{section4}

Following the detection of $\gamma$-ray flaring activity from \onees, we triggered observations at lower wavelengths, to characterize the SED of the source during the active state. The \swift\ satellite observed \onees\ on May 7, 2013 and May 8, 2013, providing spectral reconstruction in soft X-rays (using the \xrt\ telescope) and optical/UV (using the \uvot\ telescope). The \flwo\ optical telescope observed the source on May 18, 2013, ten days after the end of the \veritas\ flare. The MeV-GeV part of the SED is covered by the \fermilat\ telescope, thanks to its monitoring capabilities.\\

We performed two different analyses of \fermilat\ data: we first studied the long-term average spectrum of \onees, and then focused on the MeV-GeV emission spanning the interval of $\gamma$-ray activity detected with the \veritas\ telescopes.  Using data from August 4, 2008 to August 1, 2013, \onees\ is detected with a significance of about $14$ standard deviations over the background, and its emission between 100 MeV and 300 GeV can be parametrized by a power-law function with index  $\Gamma_{0.1-300 GeV}=1.91 \pm 0.08$ and differential flux $\Phi_{0.1-300 GeV}=(2.1 \pm 0.2)\times10^{-13}$ cm$^{-2}$ s$^{-1}$ MeV$^{-1}$, estimated at the decorrelation energy $E_{0;0.1-300 GeV}=2136$ MeV. This value is consistent with the average spectrum provided in the 2FGL catalog \cite{2FGL}. A second analysis was performed in order to produce a measurement in the GeV energy band spanning the same interval as \veritas\ observations. Only \fermilat\ observations taken between May 1 and May 7, 2013 are considered.  \onees\ is not detected by \fermilat\ in this very short period and only upper limits on its GeV emission can be computed.\\

\begin{figure}[t!]
\begin{center}
\includegraphics[width=220pt]{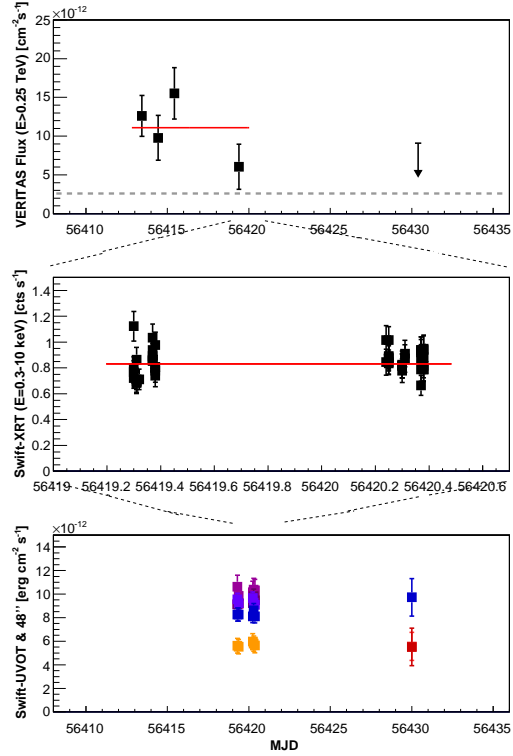}
\vspace{0.1cm}
\caption{Light curve of \onees\ during the May 2013 multiwavelength campaign. From top to bottom: the \veritas\ light-curve, the \swiftxrt\ light-curve and the \swiftuvot\ / \flwo\ light-curve. See \cite{1727apj} for details.\label{1727LC}}
\end{center}
\end{figure}

The detection of the high-flux state at VHE by \veritas\ triggered X-ray and UV observations by the \swift\ satellite, which observed \onees\ on May 7 and May 8, 2013 for a total live time of 6.8 ks. The \xrt\ light curve above $0.3$ keV (in counts per second), corrected for the exposure and for background, is shown in Fig. \ref{1727LC}. No significant variability is detected within a single observation, nor between the two observations. The average count rate measured by \xrt\ is $0.83\pm0.12$ counts per second.  Given the lack of variability, and in order to improve the statistics, the data are summed and rebinned assuming a minimum of 50 counts per bin. Data below $0.3$ keV are excluded, and the last significant spectral bin extends up to $8$ keV. The average \xrt\ spectrum is well fitted by an absorbed ($N_H=2.75\times 10^{20}$ cm$^{-2}$, as measured by \cite{Dickey90}) broken-power-law function. The best-fit parameter values are $\Gamma_1=2.01^{+0.10}_{-0.10}$, $\Gamma_2=2.44^{+0.10}_{-0.10}$, $E_{break}=1.21^{+0.24}_{-0.20}$ keV, and normalization $C=(5.3^{+0.2}_{-0.2})\times10^{-3}$ cm$^{-2}$ s$^{-1}$ keV$^{-1}$. The $\chi^2/NDF$ value is 86/84.\\

The \uvot\ telescope, on board the \swift\ satellite, observed \onees\ at optical and ultraviolet wavelengths, simultaneously with \xrt. All measurements were performed using the six available \uvot\ filters. As part of a long-term optical program of monitoring of VHE blazars, \onees\ is regularly observed by the automatic \flwo\ telescope, located near the \veritas\ site. There are no observations performed simultaneously with \veritas: the observation closest to the \veritas\ detection of \onees\ was performed on May 18, 2013, eleven days after the last detection by \veritas\ using B, r', and i' filters. The Galactic extinction is taken into account assuming $E_{B-V}=0.037$, a value consistent with the $N_H$ value used for the X-ray analysis. 
For the purpose of subtracting the host-galaxy contamination, we made use of the recent results from \cite{Nilsson07}.\\

\section{SED modeling}

\begin{figure*}[t!]
\begin{center}
\includegraphics[width=300pt]{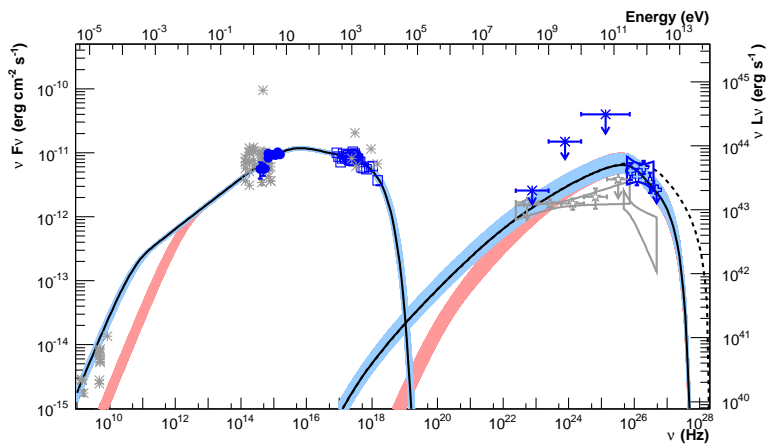}
\caption{Quasi-simultaneous SED of \onees\ (in blue; from low to high energies: \flwo, \swiftuvot, \swiftxrt, \fermilat\ and \veritas). Archival data (from the NED, \fermilat, and \magic) are included, and plotted in gray. The light-blue and pink curves represent the two sets of SSC models computed assuming $\gamma_{min}=10^3$ and $5\times10^3$. See \cite{1727apj} for further details.}
\label{figSED}
\end{center}
\end{figure*}

The SED of \onees\ is shown in Fig. \ref{figSED}. In the context of the synchrotron-self-Compton (SSC) model, the two components of the blazar SED are associated respectively with synchrotron emission from leptons (e$^\pm$) and inverse-Compton scattering of the particles off synchrotron radiation from the same lepton population. The emitting region is a spherical blob of plasma (characterized by its radius $R$) in the relativistic jet, moving towards the observer with Doppler factor $\delta$, and filled with a tangled, homogeneous magnetic field $B$. The particle population is parametrized by a broken power-law function, and it carries six free parameters: 
the minimum, maximum, and break Lorentz factors of the particles ($\gamma_{min}$, $\gamma_{max}$, $\gamma_{break}$), the two indices ($\alpha_1$ and $\alpha_2$), and the normalization factor $K$. The minimum Lorentz factor of the leptons can be fixed at a reasonably low value without affecting the modeling. The remaining eight free parameters can be constrained by observations, as discussed, for example, in \cite{Cerruti13} (and references therein).\\

 Given the lack of a simultaneous \fermilat\ detection, it is impossible to constrain the position and the luminosity of the inverse-Compton peak during the high state, and a unique solution for the SSC model cannot be provided.  However, the synchrotron component is very well sampled, and can provide some constraints  on the energy distribution of particles in the emitting region.
At low energies, the subtraction of the host-galaxy contamination reveals the AGN nonthermal continuum, which can be described by a power law from infrared to UV, with no sign of break. A fit of the \flwo\ and \uvot\ data results in an index  $n_1=1.64\pm0.09$. In the case of synchrotron radiation, this index reflects directly the index of the underlying e$^\pm$ population $\alpha_1=2n_1-1$, which is thus equal to $2.28 \pm 0.18$. Similarly, the index of the X-ray power law below the X-ray break corresponds to $\alpha_2=3.0\pm0.2$.\\

The value of $\gamma_{min}$ does not affect the modeling and it is therefore held fixed. However, values lower than $10^3$ would overestimate archival radio measurements (see Figure \ref{figSED}), even though, given their nonsimultaneity, these data should not be considered as a strong constraint. An additional constraint on $\gamma_{min}$ is provided by the \fermilat\ nondetection: in the following we study two different cases, for $\gamma_{min}=10^3$ and $5\times10^3$. The value of $\gamma_{max}$ is constrained by the break observed in the X-ray spectrum of \onees. In the following, we express it as a function of $\gamma_{break}$: $\gamma_{max}=22\ \gamma_{break}$.  \\

\label{section5}

The number of free parameters in the SSC model is five ($\delta$, $B$, $R$, $\gamma_{break}$, and $K$), and with only four observables (the frequency and flux of the synchrotron peak, the VHE spectral index and the VHE flux at 620 GeV: $\nu_{syn-peak}$; $\nu F_{\nu;syn-peak}$; $\Gamma_{VERITAS}$ and $\nu F_{\nu;VERITAS}$), a unique solution cannot be provided. However, assuming a reasonable value of the Doppler factor, it is possible to study the parameter space of the remaining free parameters. The best-fit solution is computed using the numerical algorithm described in \cite{Cerruti13}. The Doppler factor has been fixed to $30$. All the SSC models that correctly describe the \onees\ SED are then recomputed, and plotted in Fig. \ref{figSED}. The magnetic field is between 0.3 and 0.6 mG, with the emitting region size between 4$\times$10$^{17}$ and 7$\times$10$^{17}$ cm, $ and \gamma_{br}$ between 2$\times$10$^5$ and  3$\times$10$^5$. The emitting region is significantly out of equipartition ($u_e/u_B=10^{3-4}$), and the luminosity is of the order of 10$^{44}$ erg s$^{-1}$. \\

\section{Conclusions}
\label{section7}

We presented the results of a multiwavelength campaign carried out on the BL Lac object \onees\ during May 2013, triggered by a VHE high-flux state detected by \veritas. This represents the first detection of a blazar flare with \veritas\ during bright moonlight observations.\\
 
 Within this campaign, no significant variability is detected at VHE, nor at lower energies (X-rays and optical). The \veritas\ light curve is consistent with a constant flux ($6.3\%$ Crab above 250 GeV, roughly five times the archival \magic\ detection); additional observations ten days after the \veritas\ campaign indicate (at a $2.4\sigma$ level) that the high-flux state may have ended at some point after the last \veritas\ detection. 
  The quasi-simultaneous SED is fitted  by a standard one-zone SSC model. Even though the nondetection by \fermilat\ did not enable a full study of the parameter space, the measurements are fully consistent with particle acceleration by relativistic diffusive shocks and simple synchrotron and inverse Compton cooling, resulting in a power law with injection index around 2.2 and a spectral break of 1.0.\\
 
In recent years the number of VHE blazars has significantly increased thanks to the current generation of Cherenkov telescope arrays such as \veritas , \magic\ and \hess. The broadband emission and the rapid variability of blazars require prompt simultaneous multiwavelength campaigns to fully understand the physics of their emitting region. The new \veritas\ observing strategy under bright moonlight will be particularly useful for blazar science, significantly increasing blazar monitoring capabilities at VHE.\\

\end{document}